\documentclass{article}
\usepackage{spconf,amsmath,graphicx}
\usepackage{soul}
\usepackage{color}
\usepackage{multirow}
\usepackage{multicol}
\usepackage{booktabs}       % professional-quality tables
\usepackage{hyperref}
\usepackage{nicefrac}
\usepackage{stix}
\newcommand\dataset{{Melon Playlist Dataset}}

\title{Melon Playlist Dataset: a public dataset for audio-based playlist generation and music tagging}

\name{\begin{tabular}{c}Andres Ferraro$^{\star}$\quad Yuntae Kim$^{\dagger}$\quad Soohyeon Lee$^{\dagger}$\quad Biho Kim$^{\dagger}$\quad  Namjun Jo$^{\dagger}$\quad Semi Lim$^{\dagger}$\quad\\ Suyon Lim$^{\dagger}$\quad Jungtaek Jang$^{\dagger}$\quad Sehwan Kim$^{\dagger}$\quad Xavier Serra$^{\star}$\quad Dmitry Bogdanov$^{\star}$\end{tabular}}
  
  \address{$^{\star}$ Music Technology Group - Universitat Pompeu Fabra, Spain \\
      $^{\dagger}$ Kakao Corp, Korea}
\begin{document}
\sloppy
%\ninept
%
\maketitle
\begin{abstract}
One of the main limitations in the field of audio signal processing is the lack of large public datasets with audio representations and high-quality annotations due to restrictions of copyrighted commercial music. We present \dataset, a public dataset of mel-spectrograms for 649,091 tracks and 148,826 associated playlists annotated by 30,652 different tags.  All the data is gathered from Melon, a popular Korean streaming service. The dataset is suitable for music information retrieval tasks, in particular, auto-tagging and automatic playlist continuation.  Even though the latter can be addressed by collaborative filtering approaches, audio provides opportunities for research on track suggestions and building systems resistant to the cold-start problem, for which we provide a baseline. Moreover, the playlists and the annotations included in the \dataset{} %are generated by manually selected users based on the quality of the content they create, which 
make it suitable for metric learning and representation learning.

% FULL version of the abstract (190 words)

%One of the main limitations in the field of audio (music) signal processing is the lack of large public datasets with audio representations and good quality annotations due to restrictions of copyrighted commercial music. We present \dataset, a public dataset of mel-spectrograms for 649,091 tracks and 148,826 associated playlists with annotations for 30,652 different tags.  All the data is gathered from Melon, the most popular music streaming service in Korea, and is suitable for music information retrieval tasks, in particular, Automatic Playlist Continuation (APC) and auto-tagging.  Even though APC can be addressed by collaborative filtering approaches, audio provides opportunities for research on better track suggestions and building systems resistant to the cold-start problem. In this work, we provide baselines for APC in a cold-start situation. Moreover, the playlists and the annotations included in the \dataset{} are generated by manually selected users based on the quality of the content they create. The annotations of the tracks together with the playlists make it possible to apply the dataset for metric learning and representation learning, which is not possible with other public playlist datasets due to access limitations of commercial audio-content. % Well this is debatable

%The annotations of the tracks together with the playlists make the dataset suitable for methods of metric learning and representation learning that are currently not possible to apply in a public dataset. 
\end{abstract}
\section{Introduction}
\label{sec:intro}

Open access to adequately large datasets is one of the main challenges in the field of audio signal processing and music information retrieval (MIR) due to the limitations of the copyrighted material. 
%One of the main challenges in the field of audio signal processing and music information retrieval is accessing adequate and open datasets due to the limitations of the copyrighted material. 
The lack of public datasets makes collaboration between researchers and reproducibility of academic studies more difficult, limiting developments in these fields. 

In this work, we present a public dataset of information about 148,826 playlists collected by Kakao\footnote{\url{https://www.kakaocorp.com}} from Melon,\footnote{\url{https://www.melon.com}} the most popular music platform in Korean. This dataset also contains the mel-spectrogram representations of the audio for 649,091 tracks, covering the music consumed in Korea (i.e., mainly Korean pop, but also Western music). 
Thus, we provide a large-scale public dataset of playlists that includes audio information for commercial music directly accessible without the need to collect it from different external sources, which is the problem of other existing playlist datasets.
%Therefore, we provide a large-scale public dataset of playlists that includes the information from the audio, which was not provided by any public dataset before without the need of collecting it from different sources. 
The dataset can be accessed online prior registration.\footnote{\url{https://arena.kakao.com/melon_dataset}}

The playlists are collected from Melon users manually verified by moderators for providing quality public playlists. These users add metadata to the playlists, such as tags and title, which are also included in the dataset. The dataset was originally collected for the automatic playlist continuation (APC) and tag prediction challenge. Possible applications go beyond the scope of the original challenge, and the size of the dataset makes it suitable for deep learning approaches that require large amount of information. % (see Table~\ref{tab:datasets}). 
New methods can be applied for music, e.g., deep metric learning, representation learning, and semi-supervised learning.

The paper is structured as follows. %in the following way. 
We review %other 
related public datasets in Section~\ref{sec:related} and describe the proposed dataset in Section~\ref{sec:desc}.
%In Section~\ref{sec:related} we review other related public datasets. Section~\ref{sec:desc} describes the proposed dataset in detail. 
Section~\ref{sec:applications} highlights its main applications %for the dataset 
and shows an example %application
task of automatic playlist continuation %(APC) 
in a cold-start scenario. % using the audio. 
%Finally, 
Section~\ref{sec:conclusions} concludes the paper. %and discusses future work.

\section{Related work}
\label{sec:related}
Table~\ref{tab:datasets} summarizes the existing datasets for the tasks of music auto-tagging and automatic playlist continuation.

%There are some datasets that include audio and tags used in the previous years for the task of music auto-tagging but are known for their limitations. 
MagnaTagATune~\cite{law2009evaluation} (MTAT) is commonly used for auto-tagging, but mainly for prototyping because of its small size. The Million Song Dataset~\cite{bertin2011million} (MSD) %originally contained 
contains audio features extracted for one million songs, it was expanded by the MIR community with additional metadata, including collaborative tags from Last.fm. %information. %information such as lyrics and annotations. 
It was previously possible to download 30-second audio previews for MSD through the 7digital service, but it is no longer accessible.
Another limitation of this dataset is the noise in the tags~\cite{choi2018effects}. 

To address the issue of open access to audio, the FMA~\cite{defferrard2017fma} and MTG-Jamendo datasets~\cite{bogdanov2019mtg} were proposed for auto-tagging, both containing audio under Creative Commons licenses. The former is based on poorly structured music archives with inconsistent annotations and low-quality recordings. The latter tries to address this issue, focusing on a free music collection maintained for a commercial use-case, thus containing better quality audio % recordings 
and annotations. Yet, their content is different from commercial music %streaming 
platforms.

%but it contains inconsistent annotations  and low-quality recordings. Finally, the MTG-Jamendo dataset~\cite{bogdanov2019mtg} (MTG-J) was released recently and offers better quality audio recordings  Creative Commons content, which is not the most common case of commercial music platforms. 
%Table~\ref{tab:datasets} include details of these datasets. 

Recently the Million Playlist Dataset~\cite{chen2018} (MPD) was released by Spotify. This dataset contains information about one million playlists created by their  U.S. users. % located in the United States. 
However, it does not include the tracks' audio information. Even if it may be possible to download 30-second audio previews with the Spotify API, it is unclear if it is %legally allowed
legal to redistribute them. %Furthermore, 
Also, there can be inconsistencies when trying to download audio previews in the future (e.g., due to songs changing their identifier or restricted access to some of the previews in different countries). These limitations significantly affect the reproducibility and complicate the use of MPD for audio research.

%hving  it means that in the future there could be inconsistencies regarding songs that change the identifier or for some location have different access restrictions, making inconsistencies for different researchers when collecting the audio. 

The Million Playlists Songs Dataset~\cite{falcaomillion} (MPSD) combines multiple smaller datasets (Art  of  The  Mix~\cite{mcfee2012hypergraph}, \#nowplaying~\cite{pichl2015towards}, and  30Music~\cite{turrin15massimo}). Similar to MPD, this dataset does not provide audio %representations 
nor its representations for the songs.
Since it contains playlists collected from different sources, there can be noise in the data due to song matching inconsistencies between %the same song from 
multiple sources. Also, one of the source datasets, 30Music, was originally created for session-based recommendations instead of playlist continuation. % , which is a different application than playlist generation.

%Since it contains playlists collected from different sources, there can be noise in the data and mismatches between the same songs in different playlists. Also, 30Music was originally proposed for session-based recommendations, which is a different application than playlist generation.

%Therefore, %the main contribution of Kakao Dataset 
%put together
%our main contribution is 
 %information of playlists and also tags 

%After seeing 
%of the existing datasets, 

In this paper, we try to overcome the limitations of the existing datasets. Our main contribution is to provide a large research dataset of commercial music with quality playlist and tag information that includes audio representations suitable for audio-based approaches. 
Furthermore, our dataset is different because it represents music consumption in Korea instead of Western countries, bringing more cultural diversity in MIR research applied to music consumption platforms.
%Furthermore, %\dataset{} 
%our dataset is different in that it represents music consumption in Korea instead of Western countries. This is particularly important if we want to develop methods that work also in different music styles, locations, and consumption distributions. 

%In the context of the aforementioned limitations, our main contribution is to provide a large research dataset of commercial music with quality playlist and tag information that includes audio representations suitable for audio-based approaches. Furthermore, \dataset{} is different from other datasets in that it represents music consumption in Korea instead of Western countries. This is particularly important if we want to develop methods that work also in different music styles, locations, and consumption distributions. 

%Another difference in \dataset{} compared to other datasets is that represents the music consumption in Korea. 

\begin{table}[t!]
  \centering
  %\footnotesize
  \small
  \begin{tabular}{lccccc}
  \toprule
  Dataset & Tracks & Tags & Playlists & Audio (official) \\
  \midrule
  MTAT & 5,405 &188 & -- & 30 s previews \\
  MSD & 505,216 & 522,366 & -- & -- \\
  FMA & 106,574 &161 & -- & full CC tracks \\ % CC-licensed \\
  MTG-J & 55,609 & 195 & -- & full CC tracks \\
  \midrule
  MPD&  2,262,292&  --&  1,000,000&   \begin{tabular}{@{}c@{}}some previews \\ through API\end{tabular}\\
  MPSD & 1,993,607 &-- & 74,996 & -- \\
  \midrule
  %\parbox[t]{1.1mm}{\multirow{2}{*}{Kakao }} \\
  \begin{tabular}{@{}c@{}}\textbf{Melon}\\\textbf{Music}\end{tabular}&
  649,091&
  30,652&
  148,826&
  \begin{tabular}{@{}c@{}}20-50 s mel-\\spectrograms\end{tabular}\\
  \bottomrule
  \end{tabular}
  \caption{
  %The size of public datasets for automatic playlists continuation and auto-tagging and audio availability compared to \dataset{}. CC stands for audio available under Creative Commons licenses.
  Public datasets for automatic playlists continuation and auto-tagging compared to \dataset{}. CC stands for audio available under Creative Commons licenses.
  }
  \label{tab:datasets}
    
\end{table}

\section{Melon Playlist Dataset}

\label{sec:desc}

\begin{table}[ht!]
  \centering
    %\footnotesize
    \small

  \begin{tabular}{lr}
  \toprule
  Property & Count \\
  \midrule
  Track-playlist relations & 5,904,718 \\
  Unique tracks & 649,091 \\
  Tag-playlist relations & 516,405 \\
  Unique tags & 30,652 \\
  Playlists & 148,826 \\
  Playlist titles & 121,485 \\
  Unique playlist titles & 116,536 \\
  Artists & 107,824 \\
  Albums & 269,362 \\
  Genres & 30 \\
  \bottomrule
  \end{tabular}
  \caption{\dataset{} statistics.}
  \label{tab:data_stat}
\end{table}

%We originally collected all the data from Melon
All the data was originally collected from Melon
%, a Korean music streaming service, 
for a playlist continuation challenge that took place on the  Kakao Arena\footnote{\url{https://arena.kakao.com/c/8}} platform between April and July 2020  with participation of 786 teams. The dataset consists of 649,091 tracks, represented by their mel-spectrograms, and 148,826 playlists with annotations by 30,652 different tags. The playlists were created and annotated by selected users recognized for the quality of their submissions. %playlists. 
These users are named Melon DJs on the platform after Melon moderators verify them for the quality of the playlist metadata (titles, tags, and genres) they provide.

The mel-spectrograms were computed using Essentia\footnote{\url{https://essentia.upf.edu}} music audio analysis  library~\cite{bogdanov2013essentia} version \textit{2.1b5.dev677} with the following settings: 16 KHz sample rate, frame and hop size of 512 and 256 samples, and Hann window function. The scripts for their computation are provided with the dataset. %They can be reproduced using the scripts distributed with the dataset.

To reduce distributable data size, we computed mel-spectrograms only for a segment of each song (20 to 50 seconds long, not adjacent to the start or the end of the songs). Furthermore, for copyright reasons, we used a reduced 48 mel-bands resolution, which did not negatively affect the performance of the auto-tagging approaches in our previous study~\cite{ferraro2020}, while having a significantly lower reconstructed audio quality. These decisions allow saving bandwidth and disk space required to transfer and store the dataset. The dataset is distributed in 40 files, 6 GB each, with a total download size of 240 GB.
%These decisions allowed us to save bandwidth and disk space required for transfer and storage of the dataset.

%since we did not observe considerable affect on the  performance and the quality of the reconstructed audio is is low~\cite{ferraro2020}, which is a positive aspect considering to copyright limitations. In addition, by using 48 mels less bandwidth and disk space is needed to transfer and store the dataset.

%We configured Essentia to reproduce the mel-spectrograms that are generated from  another  analysis  library  used  by  the state of the art, LibROSA~\cite{mcfee2015librosa}, for compatibility. 

The dataset also includes playlist and tracks metadata. Playlist metadata contains tags and titles submitted by playlist creators, the number of users who like the playlist, and the last modification date. Track metadata contains album, title, artists, release date, and genres. The statistics of the dataset are presented in Table~\ref{tab:data_stat}. %The rate of repeated track in different playlists is 51.39\%, for the tags is 40.22\% and for the titles is  2.10\%.

\begin{figure}[ht!]
	\centering
	\includegraphics[width=0.9 \columnwidth]{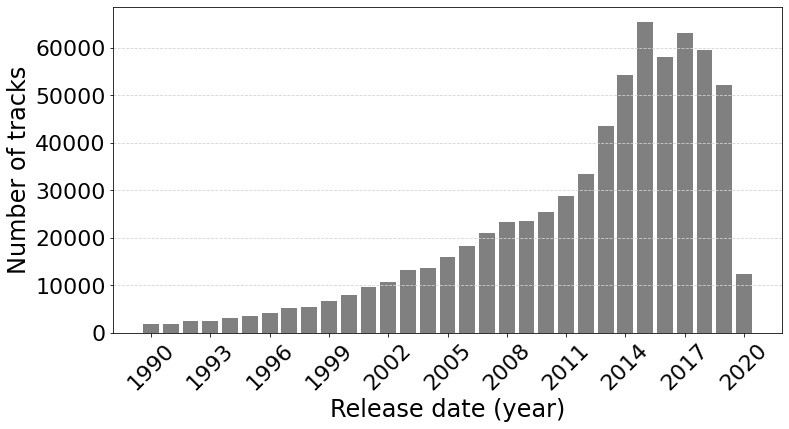}
        \caption{The distribution of release year of all tracks.}
	\label{fig:issue_date}
\end{figure}

%Regarding the year in which the tracks were published the distribution is shown in the Figure~\ref{fig:issue_date}, which indicates that 96.42\% of the tracks in the dataset were published after the year 1990.
Figure~\ref{fig:issue_date} shows the distribution of the tracks concerning their release year. Over 95\% of the tracks in the dataset were published after the year 1990.
Considering genre annotations, 25.45\% tracks in the dataset belong to only Korean music genres, 38.44\% tracks to non-Korean music genres, and 27.70\% tracks to both Korean and non-Korean genres (8.39\% tracks are annotated with music genres origin of which is unknown).

%Using the genres of the tracks, we can see that the dataset contains 25.45\% tracks annotated only with Korean music styles, 38.44\% tracks only annotated with non-Korean music styles and 27.70\% tracks are annotated with both Korean and non-Korean styles (8.39\% tracks are annotated with  unknown styles).

%The dataset is gathered from Korean music streaming service Melon for playlist continuation challenge of Kakao Arena. It consists of 148,826 playlists created by Melon DJ who is users verified by user activity history and submitted playlist quality. These playlists was seperated according to the purpose: 115,071 train (77.32\%), 23,015 validation (15.46\%), and 10,740 test (7.22\%). 
%In the dataset, there are playlist data, track meta-data, and genre code mapping table. Each row of playlist data includes its track list, tag list which tagged by a playlist creator, and also its title. Additionally, track meta-data have information such as artists, genre and issue date about all track in playlists. The statistics of the dataset is in Table~\ref{tab:data_stat}. The duplicate rate of track, tag, and title are 51.39\%, 40.22\% and 2.10\% respectively. 

\begin{figure}[ht!]
	\centering
	\includegraphics[width=0.9 \columnwidth]{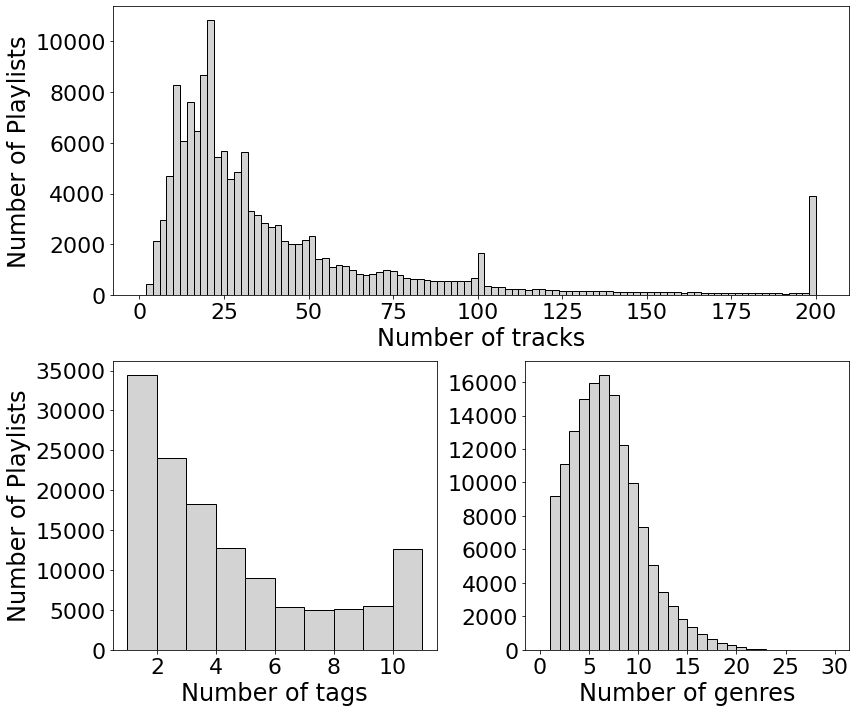}
        \caption{
            Number of tracks, tags, and genres in playlists.
            }
	\label{fig:data_dist}
\end{figure}

Playlists contain up to 200 tracks, with 41.46 tracks on average.
%The average number of tracks a playlist is 41.46 with a maximum of 200 tracks. 
The average of tags per playlist is 3.91 with a maximum of 11 tags. The number of different genres in a playlists on average is 6.31 with a maximum of 26. Figure~\ref{fig:data_dist} shows the distribution of number of tags, genres and tracks in the playlists. 

%On average the number of genres in the tracks of a playlist is 6.31 and the maximum is 26. 

\subsection{Kakao Arena challenge and the dataset split}

In the context of the challenge we divided the playlists in three groups: 115,071 playlists (77.32\%) in the train set, 23,015 playlists (15.46\%) in the validation set, and 10,740 playlists (7.22\%) in the test set. For the 33,755 playlists in validation and test sets, we considered either fully or partially hiding the tags, titles and tracks metadata. Table~\ref{tab:masked} shows the total number of playlists for each of these problem cases. %\hl{There are seven possible problem cases: 1) No track and half of others, 2) no tag and half of others, 3) no title and half of others, 4) no track, no tag, and half of title, 5) no track, no title, and half of tag, 6) no tag, no title and half of track, 7) no track, no, tag, no title.} 
The goal of the challenge was to predict the missing tracks and tags for the playlists in the test set.

%Since the challenge task is to predict tracks and tags, 33,755 playlists for evaluation and test masked a half of or all track, tag, and title.
 %Table~\ref{tab:masked} shows the number of playlists for each problem case, where $\bigtriangleup$ means half of masked data and $\bigcirc$ means no data.

Even though the challenge has finished, the Kakao Arena evaluation platform remains open for submissions of the predicted tracks and tags for the APC and auto-tagging tasks. In this way, it offers the possibility to the research community to benchmark new approaches in a standardized way using the test set with hidden tracks and tags.

%Since the challenge task is to predict tracks and tags, 33,755 playlists for evaluation and test masked a half of or all track, tag, and title. There are seven possible problems cases: 1) No track and half of others, 2) no tag and half of others, 3) no title and half of others, 4) no track, no tag, and half of title, 5) no track, no title, and half of tag, 6) no tag, no title and half of track, 7) no track, no, tag, no title. Table~\ref{tab:masked} shows the number of playlists for each problem case, where $\bigtriangleup$ means half of masked data and $\bigcirc$ means no data.

\begin{table}[ht!]
  \centering
  \small
  %\small
  \begin{tabular}{cccr}
  \toprule
  Tracks & Tags & Title & Frequency \\
  \midrule
  all & half & half & 3860 (11.43\%) \\
  half & all & half & 0 (0.00\%) \\
  half & half & all & 13165 (39.00\%) \\
  all & all & half & 2554 (7.56\%) \\
  all & half & all & 2 (0.00\%) \\
  half & all & all & 14168 (41.97\%) \\
  all & all & all & 6 (0.01\%) \\
  \bottomrule
  \end{tabular}
  \caption{Number of playlists in test and validation sets for which the tracks, tags and title were hidden either entirely (``all'') or for the half of the instances (``half'').
  %means that half of the instances were hidden and ``all'' if all the instances were hidden.
  }
  \label{tab:masked}
\end{table}

%Each playlist, which includes masked playlists, consists of 41.46 tracks on average and maximum 200 tracks in Figure~\ref{fig:data_dist}. It also has 3.91 tags on average and maximum 11 tags. The genre diversity of tracks in each playlist is included 6.31 genres on average, which categorized total 30 genres, and maximum 26 genres. Among the tracks of all playlist, 97.42\% tracks have published after 1990's, and this distribution is Figure~\ref{fig:issue_date}.

\section{Automatic Playlist Continuation}
\label{sec:applications}

\dataset{} offers many research possibilities. The most direct are playlists generation and auto-tagging for which it was originally created. %which is what the original dataset was created for. 

The task of APC consists on recommending a list of tracks to continue a given playlist. Many approaches had been proposed for this task including collaborative and content-based \cite{schedl2018current,zamani2019analysis}.
Collaborative filtering approaches usually offer the best performance according to offline metrics in the task of track recommendations to users. Given that it is not possible to recommend items without any previous interaction with these approaches (the cold-start problem), in the last years deep learning approaches have been proposed to overcome this problem %with metric learning, 
by predicting the collaborative representations from audio~\cite{liang2015content, van2013deep}.  
%In \dataset{}, for the first time a public dataset contains playlist information together with audio information of the tracks, 
%time that a 
\dataset{} is the first public dataset to contain playlist information together with directly available audio information of the tracks on a large scale,
allowing to experiment with such audio-based approaches. % for the task of APC. 
%Nevertheless, there are other applications in which this dataset can be applied.

In what follows, we provide an example of an audio-based APC approach, allowing us to expand a playlist with previously unseen tracks. We focus on underrepresented tracks in our evaluation, which is different from the Kakao Arena challenge, where the tracks in the test set had significantly more associated track-playlist interactions available for collaborative filtering. For this reason, and for reproducibility outside the Kakao Arena platform, we create an alternative split.

\subsection{Method}

We created a subset of \dataset{}, discarding the playlists with less than 5 tracks. 
For each playlist  we split its track-playlist interactions, using the tracks that appear at least in 10 playlists for our training set (\textit{APC-train}) and the rest of the tracks (considered cold-start tracks) for testing (\textit{APC-test}).
The \textit{APC-train} subset contains interactions for a total of 104,645 playlists and 81,219 tracks. 

Similar to Van den Oord et al.~\cite{van2013deep}, we train a Matrix Factorization (MF) model on the APC-train track-playlist matrix using %Weighted Approximate-Rank Pairwise 
WARP loss function~\cite{weston2011wsabie} and optimizing the parameters on 10\% of the training interactions. % APC-train-val. 

The MF model outputs the latent factors of the tracks and playlists in APC-train, we train an audio model to predict these track factors from mel-spectrograms provided in the \dataset.
%For training an audio model to predict these track factors, 
To this end, we split the tracks in APC-train into \textit{APC-train-train} (90\%) for training and \textit{APC-train-val} (10\%) for validation. 
We use a fully-convolutional neural network common for auto-tagging, based on VGGish architecture~\cite{choi2017convolutional} and trained with Mean Squared Error (MSE) as a loss function.
%In pre-evaluation 
We observed reasonable approximation of the CF track factors by the audio model, with the MSE of $0.0098$. 

Once trained, we apply the model to predict latent factors for the cold-start tracks in APC-test and match those factors to the playlist factors \cite{zamani2019analysis} in APC-train to generate rankings of the best tracks to expand those playlists. We evaluate the top-10 and top-200 rankings using MAP and nDCG \cite{ricci2011introduction} and the rest of playlist-track interactions kept as ground truth in \textit{APC-test} for the playlists.

%We could first evaluate the effectiveness of the audio model in APC-train-val. 
%with each track appearing at least in 10 playlists. %, and the tracks that are in at least 10 playlists. 
%This subset (referred as \textit{APC-train}) contains a total of 104,645 playlists and 81,219 tracks, and 
%We reserved the rest of the playlists, not contained in APC-train, for our cold-start evaluation (referred as \textit{APC-test}). 
%We use 90\% of the tracks to train the audio model and the other 10\% is used to validate the predictions, we call this subset \textit{APC-train-val}.

\begin{table}[ht!]
  \centering
  \small
  \begin{tabular}{lp{1.1cm}p{1.1cm}p{1.2cm}p{1.3cm}}
  \toprule
  Method & MAP@10&nDCG@10&MAP@200&nDCG@200 \\
  \midrule
    Random & 0.0000& 0.0001&0.0001&0.0010 \\
  Audio & 0.0159 & 0.0395&0.0135&0.0516\\
  CF& 0.0165&  0.0414&0.0148&0.0545\\
  \bottomrule
  \end{tabular}
  \caption{Performance on APC-train-val.}
  \label{tab:valid}
\end{table}

\subsection{Results}
\label{sec:results}
In all evaluations we compare the audio approach to the random baseline and the collaboration filtering approach used as our lower-bound and upper-bound baselines, respectively. 
Table~\ref{tab:valid} shows the performance on the validation set (APC-train-val). 
Comparing the performance of latent factors predicted from audio with the ones from the MF model itself,  we see that the performance of both is very similar, which shows that the audio-based approach can be used to predict latent factors for unseen tracks.

For the collaborative filtering baseline on APC-test, we use all interactions in APC-train together with 70\% of the interactions in the APC-test to train the MF model and the other 30\% to evaluate. Some test tracks are discarded from evaluation due to this split. For consistency, we use the same set of test tracks for evaluation of the rest of the approaches. 

Table~\ref{tab:test8} shows the overall performance using all considered tracks in APC-test for ranking. % and its subset with the most difficult track appearing in less than 5 playlists, APC-test-3.
In addition, we independently evaluated three subsets of APC-test described in Table~\ref{tab:test}, generating separate ranking lists among the tracks with different popularity (or ``cold-startness'') level in the dataset. %\hl{This scenario may be useful for playlist expansion with the tracks of a specific popularity level.}
The results on these subsets are given as an additional reference, but they aren't directly comparable as the performance is measured on ranking lists of different track sets. %lengths.

\begin{table}[t!]
  \centering
  \small
  \begin{tabular}{lccc}
  \toprule
  Test subset & Track in \# playlist  & Tracks& Playlists \\
  \midrule
  APC-test-1&8-9& 17,042& 27,229\\
  APC-test-2&5-8& 46,069& 35,910\\
  APC-test-3&2-5& 155,688& 31,925\\
  \bottomrule
  \end{tabular}
  \caption{Track frequency based subsets of the APC-test set.}
  \label{tab:test}
\end{table}
\begin{table}[t!]
  \small
  %\footnotesize
  %\centering
  
  \begin{tabular}{p{0.1cm}|p{0.9cm}p{1.0cm}p{1.2cm}p{1.2cm}p{1.4cm}}

  \toprule
  &Method & MAP@10&nDCG@10&MAP@200&nDCG@200 \\
  \midrule
  %\iffalse

  \multirow{3}{*}{\rotatebox[origin=c]{90}{\footnotesize{APC-test}}}  
  &Random &  0.0000& 0.0000&0.0000&0.0002\\
  &Audio &  0.0007 & 0.0014& 0.0010& 0.0052\\
  &CF &  0.0802& 0.1338&0.0581&0.1099\\
 
  \midrule  

  \multirow{3}{*}{\rotatebox[origin=c]{90}{\footnotesize{APC-test-1}}} 
  &Random & 0.0001&0.0003&0.0003&0.0022\\
  &Audio & 0.0041& 0.0065& 0.0063& 0.0267\\
  &CF & 0.0846&0.1200&0.0979&0.1923\\
  %&&&&&\\
  \midrule
  
  \multirow{3}{*}{\rotatebox[origin=c]{90}{\footnotesize{APC-test-2}}}  
  &Random & 0.0000 & 0.0000&0.0001&0.0009\\
  &Audio & 0.0022& 0.0038& 0.0032 & 0.0136\\
  &CF & 0.0490& 0.0745&0.0582&0.1291\\
  %&&&&&\\
  \midrule  
  %\fi

  \multirow{3}{*}{\rotatebox[origin=c]{90}{\footnotesize{APC-test-3}}}  
  &Random & 0.0000 & 0.0000&0.0000&0.0002\\
  &Audio & 0.0001 & 0.0001& 0.0001&0.0002\\
  &CF & 0.0274 & 0.0416&0.0341&0.0756\\
  \bottomrule
  \end{tabular}
  \caption{Performance on APC-test.
  %CF shows the upper-bound.
  }
  \label{tab:test8}
  %\vspace{-0.1cm}
\end{table}

\section{Conclusions}
\label{sec:conclusions}

%In this work 
We presented \dataset{}, the first public large-scale dataset of commercial music including the playlists, audio representation, and tags altogether, submitted by users verified for their quality annotations. 
Since the dataset reflects the music consumption in Korea, it offers %many 
novel opportunities to diversify MIR research. %in the MIR community.
%We describe %how the dataset is composed and we 
%the dataset and give an example application in which %the dataset 
%it can be used.

%More specifically, 
%We show the application of \dataset{} for playlist continuation in a cold-start scenario. 

The dataset has various applications. As an example, we considered automatic playlist continuation in a cold-start scenario and trained a baseline model to predict the latent factors of collaborative filtering from mel-spectrograms. All the code to reproduce this experiment, including the generation of dataset splits, is available online.\footnote{\url{https://github.com/andrebola/icassp2021}}

Our dataset's main limitation 
%A limitation of the dataset 
is that it provides mel-spectrograms instead of audio, making it impossible to apply methods based on other audio representations (e.g., raw waveforms). Nevertheless, the provided mel-spectrograms are suitable for the tasks of auto-tagging and automatic playlist continuation, which are the main focus of the proposed dataset. They offer a good trade-off considering the common limitations of re-using copyrighted commercial music in the field of MIR and audio signal processing. Besides, due to the large scale of the dataset, the reduced audio representations lower its distributable size, %offering advantages 
facilitating %in 
transfer and storage.

%the information that needs to be transferred and stored which o

%Given that the dataset provides the mel-spectrograms of the audio, a limitation is that the methods based on waveform representations cannot be applied. Also, there is not phase information from the audio included in the dataset which could be a limitation for some applications.  

\bibliographystyle{IEEEbib}
\bibliography{strings,refs}

\begin{thebibliography}{10}

\bibitem{law2009evaluation}
Edith Law, Kris West, Michael~I Mandel, Mert Bay, and J~Stephen Downie,
\newblock ``Evaluation of algorithms using games: The case of music tagging,''
\newblock in {\em Proc. of the 10th International Society for Music Information
  Retrieval Conference (ISMIR)}, 2009, pp. 387--392.

\bibitem{bertin2011million}
Thierry Bertin-Mahieux, Daniel~PW Ellis, Brian Whitman, and Paul Lamere,
\newblock ``The million song dataset,''
\newblock in {\em Proc. of the 12th International Society for Music Information
  Retrieval Conference (ISMIR)}, 2011.

\bibitem{choi2018effects}
Keunwoo Choi, Gy{\"o}rgy Fazekas, Kyunghyun Cho, and Mark Sandler,
\newblock ``The effects of noisy labels on deep convolutional neural networks
  for music tagging,''
\newblock {\em IEEE Transactions on Emerging Topics in Computational
  Intelligence}, vol. 2, no. 2, pp. 139--149, 2018.

\bibitem{defferrard2017fma}
Micha{\"e}l Defferrard, Kirell Benzi, Pierre Vandergheynst, and Xavier Bresson,
\newblock ``Fma: A dataset for music analysis,''
\newblock in {\em Proc. 18th International Society for Music Information
  Retrieval Conference (ISMIR)}, 2017, number CONF.

\bibitem{bogdanov2019mtg}
Dmitry Bogdanov, Minz Won, Philip Tovstogan, Alastair Porter, and Xavier Serra,
\newblock ``The mtg-jamendo dataset for automatic music tagging,''
\newblock in {\em Machine Learning for Music Discovery Workshop, International
  Conference on Machine Learning (ICML 2019)}, Long Beach, CA, United States,
  2019.

\bibitem{chen2018}
Ching-Wei Chen, Paul Lamere, Markus Schedl, and Hamed Zamani,
\newblock ``Recsys challenge 2018: Automatic music playlist continuation,''
\newblock in {\em Proc. of the 12th ACM Conference on Recommender Systems},
  2018, p. 527–528.

\bibitem{falcaomillion}
Felipe Falcao and Daniel M{\'e}lo,
\newblock ``The million playlists songs dataset: a descriptive study over
  multiple sources of user-curated playlists,''
\newblock in {\em 16th Brazilian Symposium on Computer Music}, 2017.

\bibitem{mcfee2012hypergraph}
Brian McFee and Gert~RG Lanckriet,
\newblock ``Hypergraph models of playlist dialects,''
\newblock in {\em Proc. 13th International Society for Music Information
  Retrieval Conference (ISMIR)}. Citeseer, 2012, vol.~12, pp. 343--348.

\bibitem{pichl2015towards}
Martin Pichl, Eva Zangerle, and G{\"u}nther Specht,
\newblock ``Towards a context-aware music recommendation approach: What is
  hidden in the playlist name?,''
\newblock in {\em IEEE International Conference on Data Mining Workshop
  (ICDMW)}. IEEE, 2015, pp. 1360--1365.

\bibitem{turrin15massimo}
Roberto Turrin, Massimo Quadrana, Andrea Condorelli, Roberto Pagano, and Paolo
  Cremonesi,
\newblock ``30music listening and playlists dataset,''
\newblock {\em Poster Proc. ACM Conference on Recommender Systems}, vol. 15,
  2015.

\bibitem{bogdanov2013essentia}
Dmitry Bogdanov, Nicolas Wack, Emilia G{\'o}mez~Guti{\'e}rrez, Sankalp Gulati,
  Herrera Boyer, Oscar Mayor, Gerard Roma~Trepat, Justin Salamon,
  Jos{\'e}~Ricardo Zapata~Gonz{\'a}lez, Xavier Serra, et~al.,
\newblock ``Essentia: An audio analysis library for music information
  retrieval,''
\newblock in {\em Proc. of the 14th International Society for Music Information
  Retrieval Conference (ISMIR)}, 2013.

\bibitem{ferraro2020}
Andres Ferraro, Dmitry Bogdanov, Xavier Serra, Jay~Ho Jeon, and Jason Yoon,
\newblock ``How low can you go? reducing frequency and time resolution in
  current cnn architectures for music auto-tagging,''
\newblock in {\em Proc. 28th European Signal Processing Conference (EUSIPCO)},
  2020.

\bibitem{schedl2018current}
Markus Schedl, Hamed Zamani, Ching-Wei Chen, Yashar Deldjoo, and Mehdi Elahi,
\newblock ``Current challenges and visions in music recommender systems
  research,''
\newblock {\em International Journal of Multimedia Information Retrieval}, vol.
  7, no. 2, pp. 95--116, 2018.

\bibitem{zamani2019analysis}
Hamed Zamani, Markus Schedl, Paul Lamere, and Ching-Wei Chen,
\newblock ``An analysis of approaches taken in the acm recsys challenge 2018
  for automatic music playlist continuation,''
\newblock {\em ACM Transactions on Intelligent Systems and Technology (TIST)},
  vol. 10, no. 5, pp. 1--21, 2019.

\bibitem{liang2015content}
Dawen Liang, Minshu Zhan, and Daniel~PW Ellis,
\newblock ``Content-aware collaborative music recommendation using pre-trained
  neural networks.,''
\newblock in {\em Proc. of the 16th International Society for Music Information
  Retrieval Conference (ISMIR)}, 2015, pp. 295--301.

\bibitem{van2013deep}
Aaron Van~den Oord, Sander Dieleman, and Benjamin Schrauwen,
\newblock ``Deep content-based music recommendation,''
\newblock in {\em Advances in neural information processing systems}, 2013, pp.
  2643--2651.

\bibitem{weston2011wsabie}
Jason Weston, Samy Bengio, and Nicolas Usunier,
\newblock ``Wsabie: scaling up to large vocabulary image annotation,''
\newblock in {\em Proc. of the 22nd international joint conference on
  Artificial Intelligence-Volume Volume Three}, 2011, pp. 2764--2770.

\bibitem{choi2017convolutional}
Keunwoo Choi, Gy{\"o}rgy Fazekas, Mark Sandler, and Kyunghyun Cho,
\newblock ``Convolutional recurrent neural networks for music classification,''
\newblock in {\em IEEE International Conference on Acoustics, Speech and Signal
  Processing (ICASSP)}. IEEE, 2017, pp. 2392--2396.

\bibitem{ricci2011introduction}
Francesco Ricci, Lior Rokach, and Bracha Shapira,
\newblock ``Introduction to recommender systems handbook,''
\newblock in {\em Recommender systems handbook}, pp. 1--35. Springer, 2011.

\end{thebibliography}

\end{document}